\def\o{\over}
\def\A{\rightarrow}
\def\bar{\overline}

\def\a{\alpha}
\def\b{\beta}
\def\n{\nu}
\def\m{\mu}

\def\e{\epsilon}

\def\th{\theta}

\def\Im{{\rm Im}}

\def\bar{\overline}

\def\G{{\rm GeV}}

\def\eV{{\rm eV}}

\documentstyle [12pt,epsf]{article}
\begin{document}
\baselineskip=24.5pt
\setcounter{page}{1}
\thispagestyle{empty}
\topskip 0.5  cm
\begin{flushright}
\begin{tabular}{c c}
& {\normalsize hep-ph/9612444  Revised EHU-96-12}\\
& June 1997
\end{tabular}
\end{flushright}
\vspace{1 cm}
\centerline{\Large\bf Prediction on CP Violation in Long Baseline }
\centerline{\Large\bf   Neutrino  Oscillation Experiments}
\vskip 1.5 cm
\centerline{{\bf Morimitsu TANIMOTO}
  \footnote{E-mail address: tanimoto@edserv.ed.ehime-u.ac.jp}}
\vskip 0.8 cm
 \centerline{ \it{Science Education Laboratory, Ehime University, 
 790-77 Matsuyama, JAPAN}}
\vskip 2 cm
\centerline{\bf ABSTRACT}\par
\vskip 0.5 cm
 We predict 
    $CP$ violation  in the long baseline accelerator experiments taking into consideration 
	 the  recent LSND data and the atmospheric neutrino data.
	 The estimated upper bound of  $CP$ violation is
     $0.006$, which may be observable
	 in the long baseline accelerator experiments.
	 It is found that  the upper bound increases  to $0.01$ 
	  if the LSND data is excluded.
   The matter effect, which is not $CP$ invariant,
       is found to be very small in the case we consider.
 \newpage
\section{Introduction}
  
   The origin of $CP$ violation is still an unsolved problem in particle physics.
   In the quark sector, $CP$ violation  has been intensively studied
    in the KM standard model. \cite{KM}
  For the lepton sector,  $CP$ violation is also expected,  unless
	   neutrinos are  massless.
  If neutrinos are Majorana particles, one finds three $CP$ violating phases. \cite{PH}
    However, the effect of extra Majorana phases is suppressed
	   by the factor $(m_\n/E)^2$. \cite{MJ}
    Therefore, $CP$ violation in the neutrino flavor oscillations  relates directly 
	to the $CP$ violating phase  in the mixing matrix
   for  massive neutrinos. \cite{CP}
   
   Unfortunately, this $CP$ violating effect is suppressed in the short 
 baseline neutrino oscillation
   experiments if the neurinos have a hierarchical mass spectrum.
   However,   suppression is avoidable \cite{MT,AS}
   in the long baseline accelerator experiments which are expected to operate
   in the near future. \cite{long,KEKPS} 
  Observation of  the $CP$ violating effect may be possible
    in those experiments.\par
	
    The recent indications of a deficit in the $\n_{\m}$ flux of 
   the atmospheric neutrinos \cite{Atm1}$~$\cite{Kam}
      has renewed interest in using accelerator neutrinos
    to perform  long baseline neutrino oscillation experiments.
	Many possibilities of  experiments have been discussed. \cite{long,KEKPS}
	The first experiment will begin in KEK-SuperKamiokande. \cite{KEKPS}  
	 A few authors have expressed interest in
	$CP$ violation in those experiments. \cite{MT,AS}
	We also discussed $CP$ violation  and estimated its upper bound.  
	However, we did not take into account  all available data, 
	 for example, the recent LSND data \cite{LSND} in our previous estimate.
	  By using the LSND data, we can obtain unambiguous results for $CP$ violation.
	 Thus, the purpose of this paper is to predict 
    $CP$ violation  in those accelerator experiments taking into consideration both
	 the recent LSND data \cite{LSND} and atmospheric neutrino data. \cite{Kam}
   
   On the other hand,
      the background matter effect, which is not $CP$ invariant, is a significant
     obstacle in the attempt to observe $CP$ violation originating from the phase of the neutrino
	 mixing matrix. We present  careful analyses   for  $CP$ violation in the 
	 background matter.  Therefore, we give in this paper more definite predictions compared with
	   the previous ones. \cite{MT}
	 
	 In \S 2,
	   we discuss the neutrino mass pattern, in which  
	   the recent LSND data \cite{LSND} and atmospheric neutrino data\cite{Kam}
	    are consistent with the predicted transition probabilities.
	 In \S 3, the general framework of $CP$ violation and the matter effect
	  are discussed.
  We predict the magnitude of  $CP$ violation including the matter effect
   in \S 4.  Section 5 is devoted  to  a conclusion.

 \section{Neutrino mass hierarchy and LSND data} 
 
 One expects to extract the neutrino masses and flavor mixing matrix from 
  the data of neutrino flavor  oscillations.
 Although  recent short baseline  experiments have not yet confirmed
  the evidence of  neutrino oscillation, they suggest  some
   patterns of neutrino masses and mixings.
   For example, the data  given by the LSND experiment \cite{LSND}
   indicate at least one large mass difference, such as  $\Delta m^2 \sim O(1~\eV^2)$
    with $\sin^2 2\th_{\rm LSND}  \simeq 10^{-3}$.
	In the near future, we can expect the data from the KARMEN experiment \cite{KARMEN} which
 is also searching for the $\n_\m\A\n_e(\bar \n_\m\A \bar \n_e)$ oscillation 
 as well as LSND, and the data  from the  CHORUS and  NOMAD experiments, \cite{CHONOM} which
   are looking for the $\n_\m\A\n_\tau$ oscillation. 
 The  most powerful reactor experiments searching for the neutrino oscillation
 are  those of Bugey \cite{Bugey}  and Krasnoyarsk \cite{Krasnoyarsk} at present.
They provide excluded regions in the $(\sin^2 2\th, \Delta m^2)$
 parameter space by non-observation of the neutrino oscillation.
 The first long baseline reactor experiment CHOOZ \cite{CHOOZ} will also
  give us severe constraints.
 These experiments may determine patterns of neutrino masses and mixings.
  For the present, constraints are also given by the disappearance experiments
   in CDHS \cite{CDHS} and CCFR, \cite{CCFR}  and appearance experiments
	 E531, \cite{E531} E776 \cite{E776}  and new CCFR. \cite{CCFRNEW}
 By use of those experimental bounds, some patterns 
  are studied for neutrino masses and mixings.
  \cite{Mina}$~$\cite{tani} \par
  
 In this paper, we take account of  the LSND data seriously.
 Okada and Yasuda presented clearly \cite{OYB} that
 the  Bugey reactor experiment \cite{Bugey} and the CDHS \cite{CDHS}
 accelerator experiment exclude the mass hierarchical pattern
    $ m_3\gg m_2 \gg m_1({\rm or}\simeq m_1)$ if the LSND data are used. 
 Therefore,  the mass  pattern  $m_3\simeq  m_2 \gg m_1$ 
   is favored.
   
  If the excess of the electron events at LSND are due to the neutrino oscillation, 
the $\bar \n_\m\A \bar \n_e$ oscillation
 transition probability is equal to
 $0.31^{+0.11}_{-0.10}\pm 0.05\ \%$ \cite{LSND}. 
  The plot of the LSND favored region of  $\Delta m^2$ vs $\sin^2 2\theta$  allows
    the mass region to be  larger  than $\Delta m^2\simeq 0.04\ \eV^2$.
	However, the constraint by Bugey \cite{Bugey} has  excluded
	  a mass region lower than $\Delta m^2=0.27\ \eV^2$.  On the other hand,
	the constraint by  E776 \cite{E776}  has excluded a  region
	larger than $\Delta m^2=2.3\ \eV^2$.
 Thus, the mass squared difference $\Delta m_{31}^2$ is obtained in the range
 
   \begin{equation}        
     \Delta m_{31}^2\simeq 0.27 \ \eV^2 \sim 2.3 \ \eV^2  \ .
   \end{equation}
 
  There are only two hierarchical mass difference scales 
 $\Delta m^2$  in the three-flavor mixing scheme.
 If the highest neutrino mass scale is taken to be $O(1\eV)$,
  which is appropriate for cosmological hot dark matter\ (HDM), \cite{hdm}
 the other mass scale is
  either the atmospheric neutrino mass scale  $\Delta m^2\simeq 10^{-2}\ \eV^2$
    \cite{Atm1}$~$\cite{Kam}
 or the solar neutrino mass scale $\Delta m^2\simeq 10^{-5}\sim 10^{-6}\ \eV^2$. \cite{solar}
  Since the long baseline  experiments correspond to 
  the atmospheric neutrino mass scale, 
 we take  $\Delta m^2\simeq 10^{-2}\ \eV^2$ as the lower mass scale.
 Then, one should  introduce the sterile neutrino \cite{sterile}
  if one wishs to  solve the solar neutrino problem from the standpoint of a
  neutrino oscillation.

	Since we fixed the neutrino mass scales,
	 we can now discuss the pattern of the $3\times 3$ neutrino mixing matrix.
	 In general, there are three allowed regions of the mixings, which are derived
	 from  the reactor and accelerator disappearance experiments. 
  \par
  The Bugey reactor experiment \cite{Bugey} and  CDHS \cite{CDHS}
 and CCFR \cite{CCFR}
 accelerator experiments give  bounds for the neutrino
 mixing parameters at the fixed value of $\Delta m^2_{31}$ for three cases,
 as seen in Appendix A:

\begin{equation}  
 (A)\ |U_{e 1}| \simeq 1,  \ \  |U_{\m 1}| \ll 1, \quad 
 (B)\ |U_{e 1}| \ll 1,  \ \  |U_{\m 1}| \ll 1, \quad
 (C)\ |U_{e 1}| \ll 1,  \ \ |U_{\m 1}|  \simeq 1.
 \label{ABC}
    \end{equation}
\noindent The cases (A) and (C) are   consistent with the LSND data
 because $|U_{e 1} U_{\m 1}|$ could be $O(10^{-3})$,
  as seen in Eq.\ (A$\cdot$ 6). \cite{MT,Bil,tani}
  On the other hand, the cases (A) and (B) are consistent 
  with the  atmospheric neutrino data. The following
  transition probability formulae for  the atmospheric neutrino
   may be  helpful to understand the above situation:
  \begin{eqnarray}        
P(\n_\m\A \n_\tau) &&=-4 U_{\m 2} U_{\tau 2} U_{\m 3} U_{\tau 3} 
            \sin^2({\Delta m^2_{32} L \o 4 E})+
 2 U^2_{\m 1} U^2_{\tau 1}  \ , \nonumber\\
 P(\n_\m\A \n_e) &&=-4 U_{e 2} U_{\m 2} U_{e 3} U_{\m 3} 
            \sin^2({\Delta m^2_{32} L \o 4 E})+
 2 U^2_{e 1} U^2_{\m 1}  \ .
\end{eqnarray}

 We calculate the  $CP$ violation effect in the following pattern of the
  mixing matrix (case (A)), which  is only
    consistent with  both LSND and atmospheric data:
	\begin{equation}  
 |U| \simeq  \left (\matrix{ 1 &\e_3 &\e_4 \cr 
            \e_1 &U_{\m 2}&U_{\m 3}\cr
             \e_2 &U_{\tau 2}&U_{\tau 3} \cr} \right ) \ ,
\end{equation}
\noindent
 where  $\e_i\ (i=1\sim 4)$ are small numbers.

\section{$CP$ violation and matter effect} 

 The direct measurements of $CP$ violation 
 originating from the phase of the neutrino mixing matrix are 
    the differences of the transition probabilities between $CP$-conjugate
    channels (see  Appendix B):
  \begin{eqnarray}        
  \Delta P\equiv P(\bar \n_\m \A \bar \n_e) - P(\n_\m \A \n_e) &=&
    P(\n_\m \A \n_\tau) -  P(\bar \n_\m \A \bar \n_\tau)  \nonumber \\
 = P(\bar \n_e \A \bar \n_\tau)- P(\n_e \A \n_\tau)    
   &=&  4 J^{\n}_{CP} (\sin D_{12}+\sin D_{23}+\sin D_{31}) \ ,
   \label{CPJ}
   \end{eqnarray}
  \noindent
   where  
  \begin{equation}
 D_{ij}=  \Delta m^2_{ij}{L\o 2E} \ ,
 \label{D}
   \end{equation}
   \noindent and \cite{J}
 \begin{equation}
    J_{CP}^{\n}= \Im (U_{\m 3}U^*_{\tau 3}U_{\m 2}^* U_{\tau 2})=
	 s_{12} s_{23} s_{13} c_{12} c_{23} c_{13}^2 \sin\phi \ .
   \end{equation}
   
   In the short baseline experiments, we obtain $D_{31}\simeq -D_{12} = O(1)$ 
   if $\Delta m^2_{31}=O(1\ \eV^2)$.
   The factor $(\sin D_{12}+\sin D_{23}+\sin D_{31})$ is suppressed
  because the two largest terms almost cancel due to their opposite signs.
 Another term is still small.
   Therefore,  the $CP$ violating quantity in Eq.~(\ref{CPJ}) is 
   significantly reduced.
   So one has  no chance to observe  $CP$ violation for the present
  in the short baseline neutrino oscillation experiments. 
   However, the situation is different 
   in  the long baseline accelerator experiments.
     The oscillatory terms $\sin D_{12}$ and $\sin D_{31}$  can be
	  replaced by the average value $0$
 since the magnitude of  $D_{31}$($\simeq -D_{12}$) is $10^3\sim 10^4$.
    Then $CP$ violation is dominated by the $\sin D_{23}$ term,
   which is $O(1)$.
	 Thus  the $CP$ violating quantity $\Delta P$  is not suppressed 
	  unless    $J^{\n}_{CP}$ is very small.
	\par
	In order to predict the magnitude of $CP$ violation,  one should
	 estimate $J^{\n}_{CP}$.
	In  case (A),  which is consistent with  the LSND data and the atmospheric neutrino
  data, we should take $c_{12},\  c_{13} \simeq 1$ and $c_{23}\simeq 1/\sqrt{2}$.
   The LSND data determine the value of  
  \begin{equation}
    c_{12}c_{13}|c_{23}s_{12}+s_{23}s_{13}c_{12}e^{i \phi}| \ ,
  \end{equation}
  \noindent as seen in Eq.\ (A$\cdot$ 6), if  $\Delta m^2_{31}$ is fixed.
  In the future, the observation of the process $\n_e\A \n_\tau$ 
   or $\n_\m \A \n_\tau$ will give a constraint for 
	\begin{equation}
     c_{12}c_{13} |s_{23}s_{12}-c_{23}s_{13}c_{12}e^{i \phi}| \ .
	 \label{etau}
   \end{equation}
   
   On the other hand,  the disappearance experiment in  the work by Bugey\cite{Bugey} 
   gives a bound for  $c_{12}c_{13}$ which is not so confining.
    For example, we obtain:  \cite{Bil}
    \begin{equation}
    |c_{12}c_{13}|^2 \geq 0.984 \quad {\rm at } \quad \Delta m^2_{31}=2\  \eV^2 \ .
  \end{equation}
	\noindent
	Using these constraints, we find the allowed  parameter regions of  
	$s_{12}$, $s_{13}$ and $\phi$.\par
	
	We show  the allowed curves in  the $s_{12}-s_{13}$  plane for the fixed values of $\phi$
	and $s_{23}=1/\sqrt{2}$ in Fig.~1, in which we have taken the typical
	 LSND data: $\Delta m^2_{31}=2 \eV^2$  and $\sin^2 2\th_{\rm LSND}=2.6\times 10^{-3}$. 
	 It is easily understood that larger $s_{12}$ and $s_{13}$ are allowed
	 as $\phi$ increases to $180^{\circ}$. 
	As seen in Eq.~(\ref{etau}), 
	the larger $s_{12}$ and $s_{13}$ near  $180^{\circ}$ are expected  
	  to lead to a larger transition probability of $\n_e\A \n_\tau$
   or $\n_\m \A \n_\tau$.  The predicted maximal probabilities at
   CHORUS and NOMAD are $P(\n_e \A \n_\tau)\simeq 3.3\times 10^{-4}$ and
    $P(\n_\m \A \n_\tau)\simeq 2.6\times 10^{-8}$, which are still below the
	 sensitivity of those experiments.
	  	
  It may be  useful to comment that the atmospheric neutrino anomaly is 
    attributed to the $\n_\m\A \n_\tau$ oscillation due to  $s_{23}=1/\sqrt{2}$ in this case.


   Now, we discuss the matter effect in the case of  $m_3\simeq  m_2 \gg m_1$.
 The general discussion of the matter effect in the long baseline
 experiments was given by Kuo and Pantaleone. \cite{matter}
 The $T$ violation effects in the earth were also studied numerically by
  Krastev and Petcov. \cite{matter1}
  Their results indicate that there exist relatively large regions of masses and mixings
   for which the $T$ violation effect can be considerably enhanced due to the  presence of 
   matter
  when neutrinos cross  the earth along its diameter. 
  
 Although the distance travelled by neutrinos is less than $1000 {\rm~Km}$ in
 the long baseline experiments, those data  include  the background
 matter effect which is not $CP$ invariant.  
 The matter effect should be carefully analyzed since the effect  
  depends strongly on the mass hierarchy, mixings,  and
 the incident energy of the neutrino, as shown in previous works. \cite{matter,matter1}

  The analytic formulation of the matter effect is given  for the three generation
   case in  Appendix C. \cite{matter2}
  The matter effect in the long baseline accelerater experiments
       is rather easily calculated by assuming a constant electron density, which
	    is $n_e=2.4\ {\rm  mol/cm}^3$.
However, the mixing angles and the $CP$ phase, in matter depend in a quite complicated way
  on the vacuum mass-squared differences, vacuum mixing angles and vacuum $CP$ phase
  as shown in Appendix C.
  
   We  give qualitative discussion about the matter effect by using
    the approximate formula of the  effective mixing angle $\theta_{m12}$, which is  given 
 in terms of vacuum mixings as \cite{matter}
\begin{equation}  
  \sin 2 \theta_{m12} = {\Delta m^2_{21}\sin 2\theta_{12}  \o
  \sqrt{(A \cos^2\theta_{13}- \Delta m^2_{21}\cos 2\theta_{12})^2+
  \Delta m^2_{21}\sin^2 2\theta_{12} }}  \ , \label{matt}
\end{equation}       
\noindent to zeroth order in $A \sin 2\theta_{13}$.
  In the present case of  $m_3\simeq m_2 = O(1~\eV) \gg m_1$,
  we have the relation
    $\Delta m^2_{21} \gg A\simeq 5 \times 10^{-4}~\eV^2$($A\equiv 2\sqrt{2} G_F n_e E$). 
 Then, one can easily find that the the matter effect $A$  contributes slightly
     to $\sin\theta_{m12}$ in Eq.~(\ref{matt})
  as long as $\theta_{12}\ll 1$.
 The matter effect on the  angles $\theta_{m23}$ and $\theta_{m13}$ 
   are also expected to be very small.
Thus the matter effect is expected to be small in the case we consider.
 Numerical studies are discussed in the next section.

\noindent
\section{$CP$ violation in long baseline accelerator experiments} 
    Long baseline accelerator experiments are planed to operate
   in the near future. \cite{long,KEKPS}
   The most likely possiblities are in KEK-SuperKamiokande~(250~Km),
  CERN-Gran Sasso~(730~Km) and  Fermilab-Soudan 2~(730~Km) experiments~(MINOS).
   The average energy of the $\n_\m$ beams are approximately $1~\G$,
   $6~\G$ and $10~\G$ at KEK-PS~$(12~\G)$, CERN-SPS~$(80~\G)$ and the Fermilab
	 proton accelerator $(120~\G)$, respectively.
   
       We can estimate the factor $\sin D_{12}+\sin D_{23}+\sin D_{31}$
	 in those experiments. \cite{MT}
	 Although estimated values depend on $\Delta m^2_{31}$ and $E$,    
  these  could be  close to $1$ for $\Delta m^2_{31}=1\sim 10~\eV^2$.
  In order to predict the magnitude of $CP$ violation at the  KEK-SuperKamiokande~(250~Km)
  experiment, we fix $L=250$~Km and take  $E=1.3~\G$ with $\Delta m^2_{32}=0.01~\eV^2$.
    This experiment will be a disappearance experiment,
  since such  neutrinos will not produce  tau leptons.
  But there is a manifestation of  $CP$ violation
   even in the case of disappearance  if  final states are summed over.

 Now we can calculate 
 $\Delta P(\n_\m \A \n_\tau)\equiv P(\n_\m \A \n_\tau) - P(\bar \n_\m \A \bar \n_\tau)$
  and $\Delta P(\n_\m \A \n_e)\equiv P(\n_\m \A \n_e) - P(\bar \n_\m \A \bar \n_e)$, which are
 the direct measurement of  $CP$ violation in the long baseline experiments.
  In the present case,  
 $P(\n_\m \A \n_\tau)$ is large ($\simeq 0.4$) but $P(\n_\m \A \n_e)$ is
  very small~($\simeq 0.01$).
  By use of the constraints in Fig.~1, we  predict $\Delta P(\n_\m \A \n_\tau)$
   and  $\Delta P(\n_\m \A \n_e)$ in the long baseline
   experiment with $L=250$~Km and $E=1.3 ~\G$.
   We show $\Delta P$ in the matter versus the vacuum mixing $s_{12}$
     with $\phi=159^\circ$ in Fig.~2, in which
    the predicted  $\Delta P(\n_\m \A \n_\tau)$ in the vacuum  is also presented 
	 by the dotted curve.
	The predicted  $\Delta P(\n_\m \A \n_e)$ in the vacuum cannot be distiguished from
	 that in  matter.
	It is found that the matter effect is very small.
	As $\phi$ changes from $90^\circ$ to $180^\circ$,  
	 larger $s_{12}$ and $s_{13}$ are allowed, but $\sin\phi$ decreses.
	So the maximal $J^{\n}_{CP}$ is realized between $90^\circ$ to $180^\circ$.
	Actually, the largest value of $CP$ violation is given at  $\phi=159^\circ$.
	The predicted largest value is $0.006$, which may be obserbable 
	in the KEK-SuperKamiokande experiment in the future.\par
	
 It is noticed that the $CP$ non-invariant quantity   
 $|\Delta P(\n_\m\A \n_\tau)|$ is different from $|\Delta P(\n_\m\A \n_e)|$
  in  matter.  On the other hand,
  these have the same magnitudes for  $CP$ violation originating from the
 phase in the neutrino mixing  matrix, as seen in Eq.~(\ref{CPJ}).
 

In our predictions, we have used the LSND data.
However, it may asserted that the LSND result has not yet been confirmed.
 One has to wait for  the data from  the KARMEN experiment  which
 is also searching for the $\n_\m\A\n_e(\bar \n_\m\A \bar \n_e)$ oscillation 
 as well as LSND. 
 In order to clarify the role of the constraint from the  LSND data,
  we discuss the case excluding the LSND data.
  As long as  the highest neutrino mass scale is taken to be $O(1~\eV)$,
  which is appropriate for the cosmological HDM, and 
  the atmospheric neutrino anomaly is due to  the large 
  $\n_\m\A\n_\tau$ oscillation,  
 the mass  pattern  $m_3\simeq  m_2 \gg m_1$ 
   is also favored, even if the LSND data is excluded (here, the hierarchy of masses is assumed).
   Then, the most significant constraint is given by
    the E776 data. The mass squared difference $\Delta m_{31}^2$ is  allowed
	 in the region larger than $2.3 ~\eV^2$, which is the upper bound given by
	  LSND.  In the range 
	 $\Delta m_{31}^2=2\sim 10~\eV^2$, we find
	 the largest bound of $\sin^2 2\th=3\times 10^{-3}$ at $\Delta m_{31}^2=8~\eV^2$,
	 which is not allowed by the LSND data.
 By use of this bound, we show the 
    curves of the upper bounds in  the $s_{12}-s_{13}$  plane,  fixing
	 $s_{23}=1/\sqrt{2}$ in Fig.~3, where
	the larger of $s_{12}$ and $s_{13}$ are allowed.
  The largest value of the $CP$ violation is given at  $\phi=163^\circ$.
  Predicted curves of the upper bounds 
  in the matter are  shown in Fig.~4.
	The predicted largest value is $0.01$ if the LSND data is excluded.
	Thus, the LSND data provides  a somewhat severe constraint.\par

\noindent
 \section{Conclusions}\par
 
 We have studied the direct measurements of  $CP$ violation
 originating from the phase of the neutrino mixing matrix
 in the long baseline  neutrino oscillation experiments.
 We can predict 
    $CP$ violation  in these accelerator experiments taking into consideration both
	  recent LSND data \cite{LSND} and atmospheric neutrino data. \cite{Kam}
  These data favor the mass hierarchy   $m_3\simeq  m_2 \gg m_1$ with $m_3=O(1 ~\eV)$
   and  the flavor mixings 
   $|U_{e 1}| \simeq 1$  and $|U_{\m 1}| \ll 1$ with $|U_{\m 2}| \simeq 1/\sqrt{2}$.
	The estimated upper bound of  $CP$ violation is
     $0.006$, which may be observable
	 in the long baseline accelerator experiments in the future.
   The matter effect, which is not $CP$ invariant,
       is found to be very small.
	
	In order to  clarify the role of the constraint from the  LSND data,
	 the case excluding the  LSND data  has been discussed.
	 Then, the predicted upper bound is $0.01$.
	 It is noted that the LSND data provides  the somewhat tight constraint.\par

\section*{Acknowledgements}
I would like to thank  Prof. A. Smirnov and Dr. H. Nunokawa for the  quantitative
 discussion of the matter effect. I also thank Prof. H. Minakata for  his valuable comments.
 This research is  supported by 
 the Grant-in-Aid for Science Research,
Ministry of Education, Science and Culture, Japan~(No. 07640413).

\appendix
\section{Experimental Constraints}
  We show  constraints for the neutrino mixings from  the reactor and
  accelerator disappearance experiments. 
Since no indications in favor of neutrino oscillations were found in 
these experiments, we only  obtain the  allowed regions 
in $(U^2_{\a 1}, \Delta m^2_{31})$ parameter space, where
    $\Delta m^2_{ij}\equiv m^2_i-m^2_j$ is defined.
     The Bugey reactor experiment \cite{Bugey} and  CDHS \cite{CDHS}
 and CCFR \cite{CCFR}
 accelerator experiments give  bounds for the neutrino
 mixing parameters at the fixed value of $\Delta m^2_{31}$.
 We follow the analyses given by Bilenky et al.\cite{Bil}
\par
  Since the $CP$ violating effect can be neglected 
   in those short baseline experiments, 
   we  use the following  formula without $CP$ violation
     for the probability in the disappearance experiments:
   \begin{equation}        
 P(\n_\a\A \n_\a) =1-4 |U_{\a 1}|^2(1-|U_{\a 1}|^2) 
            \sin^2({\Delta m^2_{31} L \o 4 E}) \ ,
	\end{equation}
 where  the mass hierarchy $\Delta m^2_{31}\simeq \Delta m^2_{21}\gg \Delta m^2_{32}$
  is assumed.
 The mixing parameters can be expressed
 in terms of the oscillation probabilities as \cite{Bil}
 \begin{equation}   
   |U_{\a 1}|^2={1\o 2}(1\pm \sqrt{1-B_{\n_\a\n_\a}}) \  , 
 \end{equation}   

\noindent with
 \begin{equation}   
 B_{\n_\a\n_\a}=\{1-P(\n_\a\A \n_\a)\}\sin^{-2}({\Delta m^2_{31} L \o 4 E})\ ,  
 \end{equation}
    
\noindent where $\a=e$ or $\m$.
Therefore the parameters $U_{\a 1}^2$ at the fixed value of 
$\Delta m^2_{31}$ should satisfy one of the following inequalities: 
 \begin{equation}          
  |U_{\a 1}|^2 \geq {1\o 2}(1 + \sqrt{1-B_{\n_\a\n_\a}})\equiv a_\a^{(+)} \ ,
    \quad {\rm or} \quad  \label{cond}
 |U_{\a 1}|^2 \leq {1\o 2}(1 - \sqrt{1-B_{\n_\a\n_\a}})\equiv a_\a^{(-)} \ . 
  \end{equation}   
\noindent The negative results of  Bugey, CDHS  
  and CCFR experiments have given the values of $a_e^{(\pm)}$ and
$a_\m^{(\pm)}$. \cite{Bil}
For example, one obtains 
 $a_e^{(+)}=0.984$,  $a_e^{(-)}=0.016$,  $a_\m^{(+)}=0.985$ and  $a_\m^{(-)}=0.015$
  for $\Delta m^2_{31}=2 ~\eV^2$.
\par
 It is noted from Eq.~(\ref{cond}) there are three allowed regions of 
$|U_{e 1}|^2$
 and $|U_{\m 1}|^2$ as follows:
\begin{eqnarray}  
 (A)\quad |U_{e 1}|^2 \geq a_e^{(+)} \ , \qquad  |U_{\m 1}|^2 \leq  
 a_\m^{(-)} \ ,
    \nonumber \\
 (B)\quad |U_{e 1}|^2 \leq a_e^{(-)} \ , \qquad |U_{\m 1}|^2 \leq  
 a_\m^{(-)} \ ,
      \label{ABC} \\     
 (C) \quad |U_{e 1}|^2 \leq a_e^{(-)} \ , \qquad |U_{\m 1}|^2 \geq  a_\m^{(+)}
 \ .
    \nonumber
\end{eqnarray}
\noindent
 In addition to these constraints,  we should take account of
 the constraints of  E531 \cite{E531} and E776 \cite{E776} experimental data
 by using the  transition probabilities
  \begin{eqnarray}   
     P(\n_\m\A \n_e) \simeq  4 |U_{e 1}|^2 |U_{\m 1}|^2 
            \sin^2({\Delta m^2_{31} L \o 4 E}) \ ,  \nonumber \\
     P(\n_\m\A \n_\tau) \simeq 4 |U_{\m 1}|^2 |U_{\tau 1}|^2 
            \sin^2({\Delta m^2_{31} L \o 4 E}) \ .
\end{eqnarray} 
 These constraints often become  severer than
  the ones of the disappearance experiments.

\section{$CP$ Violation}
  The amplitude of the  $\n_\a\A \n_\b$ transition with
	 the neutrino  energy $E$
   after traversing the distance $L$ can be written as
  \begin{equation}        
 {\cal{A}}(\n_\a\A \n_\b) = e^{-i EL} \left\{\delta_{\a\b} 
   + \sum_{k=2}^3  
 U_{\a k} U_{\b k}^* \left [\exp{\left (-i {\Delta m^2_{k1} L \o 2 E}\right )}
	   -1 \right ] \right\} \ , 
		 \label{Pro}
\end{equation}
\noindent  where   
the $U_{\a i}$ denote the elements of the $3\times 3$ neutrino  flavor 
mixing matrix, in which $\a$ and $i$  refer to the flavor eigenstate and 
 the mass eigenstate, respectively. 
  The  amplitude  ${\cal A} (\bar\n_\a\A \bar\n_\b)$ is given by 
  replacing $U$ with $U^*$ on the right-hand side of Eq.~(\ref{Pro}).
 The direct measurements of $CP$ violation 
 originating from the phase of the neutrino mixing matrix are 
    the differences of the transition probabilities between $CP$-conjugate
    channels:\cite{CP}  
  \begin{eqnarray}        
  \Delta P &&\equiv P(\bar \n_\m \A \bar \n_e) - P(\n_\m \A \n_e) =
    P(\n_\m \A \n_\tau) -  P(\bar \n_\m \A \bar \n_\tau)  \nonumber \\
 &&= P(\bar \n_e \A \bar \n_\tau)- P(\n_e \A \n_\tau)    
   =  4 J^{\n}_{CP} (\sin D_{12}+\sin D_{23}+\sin D_{31})\ , 
   \label{CP}
   \end{eqnarray}
  \noindent
   where  
  \begin{equation}
 D_{ij}=  \Delta m^2_{ij}{L\o 2E} \ ,
 \label{D}
   \end{equation}
 and
 $J^{\n}_{CP}$ is defined for the  rephasing  invariant quantity of  
 $CP$ violation   in the neutrino mixing  matrix as well as that 
in the quark sector.\cite{J}
 In terms of the standard parametrization of the  mixing matrix,\cite{PDG}
 
\begin{equation}  
  \left (\matrix{ c_{13} c_{12} & c_{13} s_{12} &  s_{13} e^{-i \phi}\cr 
  -c_{23}s_{12}-s_{23}s_{13}c_{12}e^{i \phi} & c_{23}c_{12}-s_{23}s_{13}s_{12}e^{i \phi} & 
                       s_{23}c_{13} \cr
  s_{23}s_{12}-c_{23}s_{13}c_{12}e^{i \phi} & -s_{23}c_{12}-c_{23}s_{13}s_{12}e^{i \phi} & 
                       c_{23}c_{13} \cr} \right ) \ ,
\end{equation}

\noindent   where  $s_{ij}\equiv \sin{\theta_{ij}}$ and $c_{ij}\equiv \cos{\theta_{ij}}$
 are vacuum mixings, and $\phi$ is the $CP$ violating phase,   we have
	 \begin{equation}
    J_{CP}^{\n}= \Im (U_{\m 3}U^*_{\tau 3}U_{\m 2}^* U_{\tau 2})=
	 s_{12} s_{23} s_{13} c_{12} c_{23} c_{13}^2 \sin\phi \ .
	\end{equation}
   \noindent  The oscillatory terms are periodic in $L/E$, and  
   $D_{12}+D_{23}+ D_{31}=0$ is satisfied.

\section{Matter Effect}
 The effective mass squared in matter $M_m^2$ for the
 neutrino energy $E$ in the weak basis \cite{matter2} is

\begin{equation}  
 {M_m^2} = U_m \left (\matrix{ m_1^2 & 0 & 0 \cr 
            0 & m_2^2 & 0 \cr
            0 & 0 & m_3^2 \cr} \right )U_m^\dagger +
      \left (\matrix{ A & 0 & 0 \cr 
            0 & 0 & 0 \cr
            0 & 0 & 0 \cr} \right )  \ , 
			\label{mass}
\end{equation}
\noindent   
where $A\equiv 2\sqrt{2} G_F n_e E$.
 For antineutrinos, the effective mass squared is given by
 replacing $A\A -A$ and $U_m \A U_m^*$. 
 The effective mixing matrix $U_m$ is
\begin{equation}  
  \left (\matrix{ c_{m13} c_{m12} & c_{m13} s_{m12} &  s_{m13} e^{-i \phi_m} \cr 
  -c_{m23}s_{m12}-s_{m23}s_{m13}c_{m12} e^{i \phi_m} 
  & c_{m23}c_{m12}-s_{m23}s_{m13}s_{m12}e^{i \phi_m} & s_{m23}c_{m13} \cr
  s_{m23}s_{m12}-c_{m23}s_{m13}c_{m12} e^{i \phi_m} &
    -s_{m23}c_{m12}-c_{m23}s_{m13}s_{m12} e^{i \phi_m} &  c_{m23}c_{m13} \cr} \right ) 
\end{equation}       
\noindent
 where $s_{mij}\equiv \sin{\theta_{mij}}$ and $c_{mij}\equiv \cos{\theta_{mij}}$
 are  effective mixings,  and $\phi_m$ is the effective phase in  matter.
 
  If  we use the constant
 electron density $n_e=2.4\ {\rm  mol/cm}^3$, the effective mixing angles and phase are given 
 in terms of vacuum mixings and the effective neutrino masses  $M_1$, $M_2$ and $M_3$,
  which are eigenvalues in Eq.~(\ref{mass}),  as follows: \cite{matter2}

\begin{eqnarray}  
  s_{m12}^2 &=& {-(M_2^4-\a M_2^2+\b) \Delta M^2_{31} \o
  \Delta M^2_{32}(M_1^4-\a M_1^2+\b)-\Delta M^2_{31}(M_2^4-\a M_2^2+\b)} \ , \nonumber \\
    s_{m13}^2 &=& {M_3^4-\a M_3^2+\b \o \Delta M^2_{31} \Delta M^2_{32}} \ , \nonumber  \\
 s_{m23}^2 &=& {E^2 s^2_{23} + F^2 c^2_{23} + 2 E F c_{23}s_{23}\cos\phi\o
          E^2 + F^2} \ ,   \\
  e^{-i \phi_m} &=& {(E^2 e^{-i\phi} - F^2 e^{i\phi}) s_{23}c_{23} + 
   E F(c^2_{23}-s^2_{23}) \o
          \sqrt{(E^2 s^2_{23} + F^2 c^2_{23} + 2 E F c_{23}s_{23}\cos\phi)
		   (E^2 s^2_{23} + F^2 c^2_{23} - 2 E F c_{23}s_{23}\cos\phi )}} \ , \nonumber 
\end{eqnarray} 
\noindent where
\begin{eqnarray}  
 \a &=& m_3^2 c_{13}^2 + m_2^2(c^2_{12}c^2_{13}+ s^2_{13})+ m_1^2(s^2_{12}c^2_{13}+ s^2_{13}) \ ,
  \nonumber \\
\b &=& m_3^2 c_{13}^2 (m_2^2 c^2_{12}+ m_1^2 s^2_{12})+m_2^2 m_1^2 s^2_{13} \ , \nonumber \\
 E &=& [\Delta m^2_{31}(M_3^2- m_2^2)-\Delta m^2_{21}(M_3^2- m_3^2)s^2_{12}] c_{13}s_{13} 
   \ , \nonumber \\
 F &=& (M_3^2- m_3^2)\Delta m^2_{21}c_{12} s_{12}c_{13} \ , 
\end{eqnarray} 
\noindent  and
\begin{eqnarray}  
 M^2_1&\simeq & {1\o 2}[m_1^2+m_2^2+A c_{13}^2 - \sqrt{(A c_{13}^2-\Delta m_{21}^2
   \cos 2\th_{12})^2+(\Delta m_{21}^2\sin 2\th_{12})^2} ]  \ ,\nonumber \\
 M^2_2&\simeq & {1\o 2}[m_1^2+m_2^2+A c_{13}^2 + \sqrt{(A c_{13}^2-\Delta m_{21}^2
   \cos 2\th_{12})^2+(\Delta m_{21}^2\sin 2\th_{12})^2} ]    \ ,  \nonumber \\
 M^2_3&\simeq &  m_3^2 + A s_{13}^2   \  . 
 \end{eqnarray}


\newpage
\begin{figure}
 \epsfxsize=14 cm
 \centerline{\epsfbox{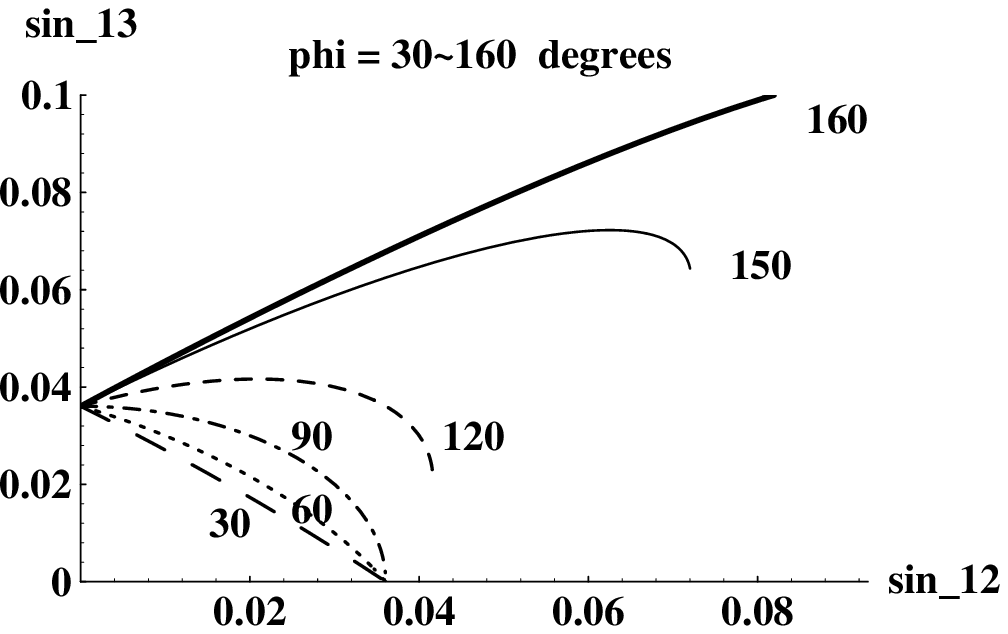}}
 \caption{}
	 \end{figure}
 Allowed curves in the $s_{12}-s_{13}$  plane,  fixing
	 $s_{23}=1/\sqrt{2}$, in the case   $\Delta m^2_{31}=2 ~\eV^2$ 
	  and $\sin^2 2\th_{\rm LSND}=2.6\times 10^{-3}$. 
	 The thick-solid curve, thin-solid curve,  short-dashed curve, dash-dotted curve, dotted curve
	  and long-dashed curve corrspond to
   $\phi$=$160^{\circ}$, $150^{\circ}$, $120^{\circ}$, $90^{\circ}$, $60^{\circ}$ and $30^{\circ}$,
	    respectively.
\newpage

\begin{figure}
 \epsfxsize=14 cm
 \centerline{\epsfbox{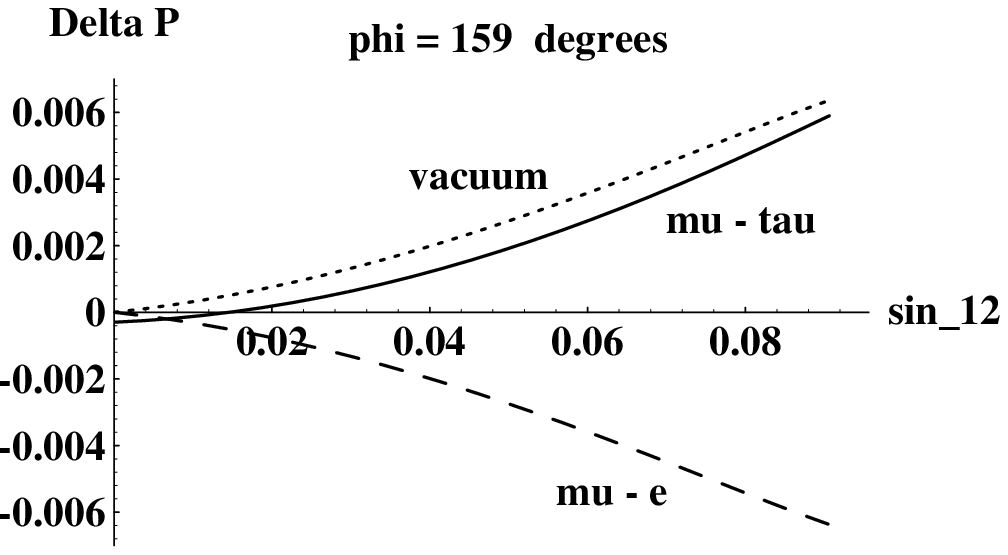}}
 \caption{}
\end{figure}
Predicted  $\Delta P(\n_\m \A \n_\tau)$(solid curve) and 
 $\Delta P(\n_\m \A \n_e)$(dashed curve) 
  in the matter versus $s_{12}$  with $\phi=159^\circ$.
      $\Delta P(\n_\m \A \n_\tau)$ in the vacuum  are also presented
	   by the dotted curve.
	 Here, $E=1.3 ~\G$ and $L=250$~Km with  $\Delta m^2_{32}= 10^{-2}~\eV^2$ are used.
\newpage
\begin{figure}
 \epsfxsize=14 cm
 \centerline{\epsfbox{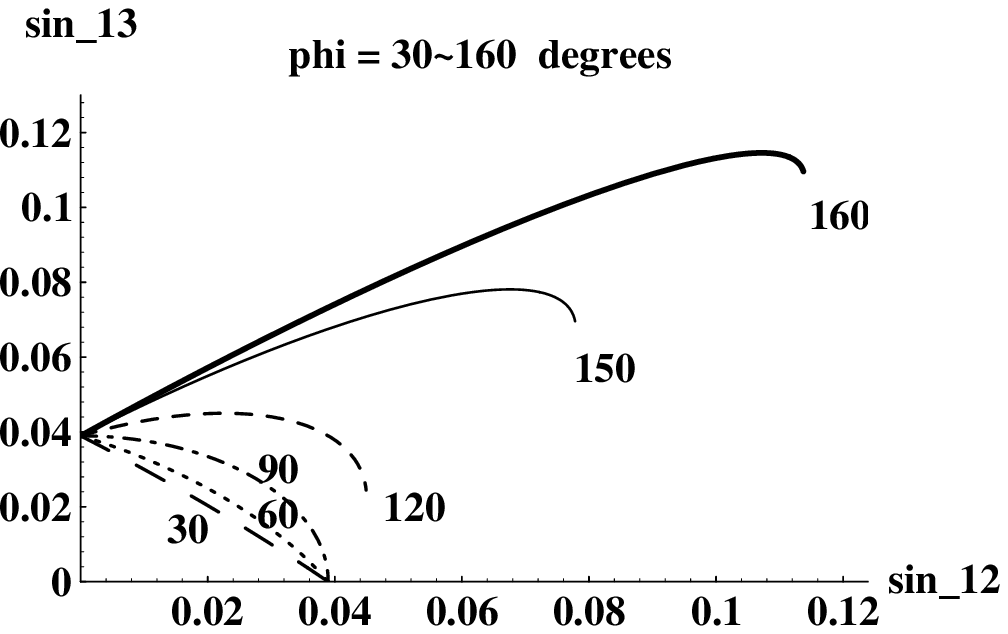}}
 \caption{}
	 \end{figure}
 Upper bounds in the  $s_{12}-s_{13}$  plane,  fixing
	 $s_{23}=1/\sqrt{2}$, in the case  $\Delta m^2_{31}=8 ~\eV^2$ 
	  and $\sin^2 2\th_{\rm E776}=3.0\times 10^{-3}$. 
	The notation  is the same as that in Fig.~1.
\newpage

\begin{figure}
 \epsfxsize=14 cm
 \centerline{\epsfbox{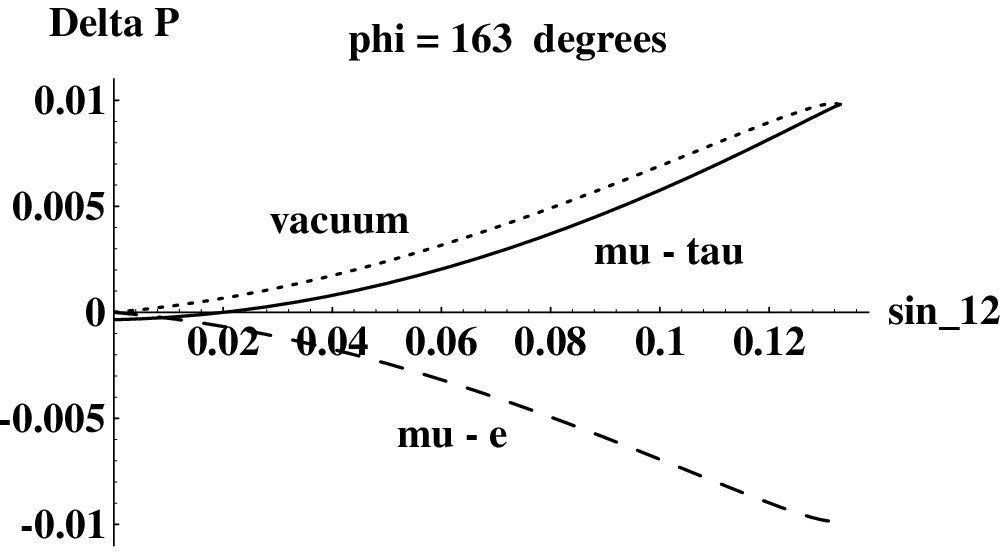}}
 \caption{}
\end{figure}
Predicted upper bound of $CP$ violation
  in  matter versus $s_{12}$  with $\phi=163^\circ$.
 The notation is the same as that  in Fig.~2.
\end{document}